\documentclass[prl,twocolumn,superscriptaddress,showpacs,amsmath,amssymb]{revtex4-1}

\usepackage{graphicx}
\usepackage{latexsym}
\usepackage{amsmath}
\usepackage{amssymb}
\usepackage{amsfonts}
\usepackage{changes}
\usepackage{color}
\usepackage{bm}
\usepackage{verbatim}
\usepackage{pagecolor,lipsum}
\usepackage{lineno}
\usepackage[version=3]{mhchem}

\definecolor{MS-color}{rgb}{0,0.0,1.0}

\usepackage{framed}
\definecolor{shadecolor}{RGB}{222,222,221}
\begin{document}

\title{Dynamic spin-triplet order induced by alternating electric fields in superconductor-ferromagnet-superconductor Josephson junctions}

\author{I. V. Bobkova}
\affiliation{Institute of Solid State Physics, Chernogolovka, Moscow
  reg., 142432 Russia}
\affiliation{Moscow Institute of Physics and Technology, Dolgoprudny, 141700 Russia}
\affiliation{National Research University Higher School of Economics, Moscow, 101000 Russia}

\author{A. M. Bobkov}
\affiliation{Institute of Solid State Physics, Chernogolovka, Moscow reg., 142432 Russia}

\author{M.A.~Silaev}
\affiliation{Department of
Physics, Nanoscience Center, University of Jyv\"askyl\"a, 40014 Jyv\"askyl\"a, Finland}
\affiliation{Moscow Institute of Physics and Technology, Dolgoprudny, 141700 Russia}
\affiliation{Institute for Physics of Microstructures, Russian Academy of Sciences, 603950 Nizhny Novgorod, GSP-105, Russia}

\date{\today}


\begin{abstract}
Dynamic states offer extended possibilities to control  the properties of quantum matter.
Recent efforts are focused on studying the ordered states which appear exclusively under the time-dependent drives. Here we demonstrate a class of systems which feature dynamic spin-triplet superconducting order stimulated by the alternating electric field. 
    The effect is based on the 
  interplay of ferromagnetism, interfacial spin-orbital coupling (SOC) and 
  the condensate motion driven by the field, which converts hidden static p-wave order, produced by the joint action of the ferromagnetism and the SOC, into dynamical s-wave equal-spin triplet correlations.  We demonstrate that  the critical current of Josephson junctions hosting these states is proportional to the  electromagnetic power, supplied either by the external irradiation or by the ac current source.
  Based on these unusual properties we propose the scheme of  a
  Josephson transistor which can be switched by the ac voltage and   demonstrates an even-numbered sequence of Shapiro steps. %
  Combining the photo-active Josephson junctions with recently discovered Josephson phase batteries we find  photo-magnetic SQUID devices which can generate spontaneous magnetic fields while being exposed to irradiation.
\end{abstract}

  \maketitle

 
Weak links between two superconducting electrodes 
known as the Josephson junctions (JJ) are the 
cornerstone elements of superconducting electronics.
For decades there has been an intensive search of technologies and physical principles allowing for the construction of   superconducting transistors based on the JJ circuits with controllable switching between superconducting and resistive states \cite{clark1980feasibility}. 
 Such devices 
 are expected to pave the way for energy-saving 
  superconducting computers\cite{likharev1991rsfq}.
Recently the interest to JJs with electrically-tunable critical currents has been stimulated by the perspectives of applying such systems in leading-edge quantum information architectures \cite{larsen2015semiconductor,
casparis2016gatemon}. 
 Main efforts  in this field have been focused on the systems with  Josephson currents controlled by electrostatic gates. 
 This concept
  has been realized in mesoscopic systems with normal metal interlayers \cite{volkov1995new,morpurgo1998hot,baselmans1999reversing,huang2002observation, paolucci2019field,de2018metallic}, semiconducting interlayers \cite{clark1980feasibility,doh2005tunable,abay2014charge, larsen2015semiconductor,
casparis2016gatemon} and   quantum dots  \cite{van2006supercurrent, szombati2016josephson}.  
Electrostatic control with constant gate voltages is not enough for most of the applications implying transistors operating under the action of high-frequency drives.
Therefore it is of crucial importance to go beyond the electrostatic gating and  find the physical mechanisms  which could provide a dynamical switching  of Josephson junctions by application of a high-frequency electric field.

 Here we suggest a qualitatively different  way to controlling the Josephson current using  dynamic triplet superconducting states driven by the external time-dependent electric field $\bm E(t)$. 
 This mechanism can help to achieve switching rates in the teraherz and even the visible light frequency domains.  It is based on the peculiar quantum state of matter which arises under the non-equilibrium conditions due to the 
 interplay of Rashba-type 
 \cite{ast2007giant,bihlmayer2006rashba,manchon2015new}
interfacial spin-orbital coupling (SOC), ferromagnetism and oscillating motion of Cooper pairs driven by the alternating electric field. The first two ingredients acting together provide partial conversion of singlet correlations to p-wave equal-spin triplet correlations, which do not manifest themselves in the diffusive system due to impurity averaging. The last ingredient converts these "hidden" static p-wave to dynamic s-wave equal-spin triplet correlations via the Doppler shift mechanism. 
The {\it triplet } nature of the proposed light-induced dynamical correlations provides an additional advantage opening a perspective of photon-magnon coupling mediated by the triplet correlations.  The proposed effect extends the possibilities of  generating
and controlling dynamical and non-equilibrium states of matter which have attracted significant attention recently, such as Floquet topological
insulators \cite{lindner2011floquet}, odd-frequency superconductivity \cite{Cayao2021}, time crystals
\cite{wilczek2013superfluidity,yao2017discrete,autti2018observation},
driven Dirac materials\cite{triola2017excitonic,Hubener2017,Sentef2015}, 
light-induced and light-manipulated superconductivity
\cite{fausti2011light,claassen2019universal,Tindall2020,Budden2021,Mitrano2016,Sentef2016,Yu2021}, vortex states\cite{yokoyama2020creation,mironov2021inverse}, cavity-enhanced ferroelectric phase transition \cite{Ashida2020}
and dynamical hidden orders \cite{balatsky2020quantum,linder2019odd,aeppli2020hidden, stojchevska2014ultrafast}.


\begin{figure}
 \centerline{$
 \begin{array}{c}
 \includegraphics[width=3.5in]
 {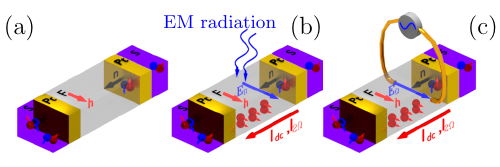} 
 \end{array}$}
 \caption{\label{Fig:1} 
   {\bf Schematic picture of the system considered.}
   S/F junction with Rashba SOC at the interface induced by the thin layer of heavy metal Pt. (a) Only short-range superconducting correlations are present shown by the blue and red spheres with opposite arrows.   (b,c) generation of long-range triplet (LRT) correlations due to the irradiation of the setup with electromagnetic wave (b) and 
   by applying the the ac current source (c) both producing the electric field $\bm E (t)= \bm E_\Omega e^{i\Omega t}$ in the ferromagnetic interlayer. The LRT are shown schematically by the red spheres with co-directed arrows corresponding to the spin states aligned with the exchange field $\bm h$. 
   }
 \end{figure}
 
  Up to now the external control of spin-triplet superconductivity has been considered mostly with the help of static fields while several works have studied the effect of magnetization precession on the Josephson current 
  \cite{zhu2004novel, houzet2008ferromagnetic, PhysRevLett.103.037003,
PhysRevB.83.104521, yokoyama2009tuning}.    
 In general, the spin-triplet pairing amplitude $\hat f$ 
   can be written in terms of the spin vector\cite{leggett1975theoretical} $\bm d =(d_x, d_y, d_z)$  
   
 \begin{align} \label{Eq:PairingAmplitude}
     \hat f = (d_x - i d_y) |\uparrow \uparrow\rangle + (d_x + i d_y) |\downarrow \downarrow\rangle + d_z (|\uparrow \downarrow \rangle + | \downarrow \uparrow \rangle )
 \end{align}
 In this paper we consider an S/F/S Josephson junction sketched in Fig.~\ref{Fig:1} with Rashba SOC at the S/F interfaces and demonstrate that external time-dependent electric field produces triplet correlations with the energy and time-dependent spin vector constructed as follows 
\begin{align}\label{Eq:SpinVector}
     \bm d (\varepsilon,t) = \int dt^\prime K_d (\varepsilon,t-t^\prime) (\bm E (t^\prime) \times \bm n) \times \bm h,
 \end{align}
 where $\bm n$ is a normal to the interface plane with  Rashba SOC. The scalar kernel $K_d (\varepsilon, t-t')$ is determined below in the framework of a microscopic model. The spin vector $\bm d$ in Eq.~(\ref{Eq:SpinVector}) is perpendicular to the exchange field $\bm h $ of the ferromagnet. Therefore, according to Eq.~(\ref{Eq:PairingAmplitude}) it describes  superconducting correlations characterized by the spin projections $\pm 1$ on the direction of the exchange field. 
 This shows up through the property of such pairs to be robust to the spin depairing. 
 At the distances $x\gg \xi_F$ 
only such pairs can survive in the ferromagnet hence named long-range triplets  (LRT). Here $\xi_F$ is the coherence length for opposite-spin pairs in the ferromagnet.
 In the absence of the electric field 
 only the short-range pairs, which are localized at the coherence length 
 $\xi_F  \sim 1$ nm near the superconducting electrodes, are  produced, 
  as shown schematically  in Fig.\ref{Fig:1}.  
Therefore we find the mechanism of electrically  stimulated spin-triplet superconductivity, which can support the long-range Josephson current through thick F layer as shown in Fig.\ref{Fig:1}b,c. We suggest that such system can be considered as the photo-active Josephson junction (JJ). This terminology means that the Josephson current is switched on by the alternating electric field originating e.g. from the  external electromagnetic radiation, as in Fig.\ref{Fig:1}b. 

Our quantitative calculations are based on the non-stationary version of Usadel-Keldysh theory of superconductivity. The main quantity entering the theory is the pairing amplitude $\check f^{R,A}= \hat f_s^{R,A} + { \bm d^{R,A}} \bm\sigma$, which can be written as the sum of spin-singlet $\hat f_s^{R,A}$ and spin-triplet $ {\bm d}^{R,A}\bm \sigma$ components. The pairing amplitude  in the ferromagnetic part of the structure is described by the linearized Usadel equation 
 \begin{eqnarray}  \label{Eq:Usadel_linearized_1}
 \pm i D \partial_x^2 \check f^{R,A} = 2 \varepsilon \check f^{R,A} - \{ \bm h \hat {\bm \sigma}, \check f^{R,A} \},
 \end{eqnarray}
  where $D$ is the diffusion constant, the $\pm $-sign refers to the retarded (R) and advanced (A) components of the pairing amplitude. The linearized theory is only valid for the weak proximity effect. In our calculation the weakness of the proximity effect is justified by the condition that temperature $T$ is close to the critical temperature $T_c$. The alternating electric field $\bm E (t) = \sum \limits_i \bm E_{\Omega_i} e^{i \Omega_i t} $ is described by the 
 time-dependent vector potential 
 $ \bm E = -  \partial_t \bm A /c $.  We assume that  $p_{s,i} \xi_{\Omega_i} \ll 1$ and $p_{s,i} \xi_S \ll 1$, where $p_{s,i} = 2eE_{\Omega_i}/\Omega_i$ is the absolute value of the condensate momentum at a given frequency, $\xi_S = \sqrt{D/\Delta}$ and $\xi_{\Omega_i} = \sqrt{D/\Omega_i}$ are the coherence lengths in the superconductor and the nonsuperconducting metal, respectively.  Then the small terms $\propto \bm A_{\Omega_i} \bm A_{\Omega_j}$, which are not spin-active and do not lead to any singlet-triplet or SRT-LRT conversion, are neglected in Eq.~(\ref{Eq:Usadel_linearized_1}). 
 For the spin- singlet component there is a usual Kupriyanov-Lukichev boundary condition \cite{Kuprianov1988}
  \begin{eqnarray}
   (\bm n \bm \nabla) \hat f_s^{R,A} =  \gamma \hat F_{bcs}^{R,A},
      \label{bc_singlet}
  \end{eqnarray}
  where $\gamma$ is the S/F interface conductance, $\hat F_{bcs}^{R,A} = \mp \tau_3 \hat \Delta /(\varepsilon \pm i \delta)$
  and $\hat \Delta = |\Delta(x)|exp[i \chi (x) \hat \tau_3]\hat \tau_1$. We assume $|\Delta(x)| = 0$ in the interlayer of the Josephson junction $-d_F/2 \leq x \leq d_F/2$, while $|\Delta(x)| = \Delta $ and $\chi(x) = \mp \chi/2$ is the superconducting phase in the left (right) leads. $\hat \tau_i$ and $\hat \sigma_i$ are Pauli matrices in particle-hole and spin spaces, respectively, and $\hat {\bm \sigma} = (\hat \sigma_1, \hat \sigma_2, \hat \sigma_3)^T$. 

   The presence of Pt layer is modelled by the 
  Rashba constant $\alpha(x)$ which is only nonzero in the restricted region near the S/F interface.
    We introduce $\tilde{\alpha} = \int dx \alpha (x)$ as the surface SOC strength and obtain
   the boundary condition for the spin-triplet component \cite{supplementary}, which is our first main result
   \begin{eqnarray} 
  (\bm n \bm \nabla) \bm d^{R,A} = \frac{4ie}{c} \tilde \alpha \hat \tau_3 \sum \limits_i e^{i \Omega_i t} \bigl( \bm A_{\Omega_i} \times \bm n \bigr) \times \nonumber \\
  \bigl[ \bm d^{R,A}(\varepsilon + \frac{\Omega_i}{2})+\bm d^{R,A}(\varepsilon - \frac{\Omega_i}{2}) \bigr].
   \label{bc_d}
  \end{eqnarray}

  In general the solution of Eq.~(\ref{Eq:Usadel_linearized_1})  consists of short-range and long-range modes. They decay in the ferromagnetic region at the distances of $\xi_F = \sqrt{D/h}$ and $\xi_{\Omega_i} $, respectively. Solving  Eqs.~(\ref{Eq:Usadel_linearized_1}), (\ref{bc_singlet}),
  (\ref{bc_d}) for the Josephson setups Fig.\ref{Fig:1}b,c  we obtain\cite{supplementary} the LRT:
 \begin{eqnarray}
  \bm d^{R,A} (\varepsilon,t) = \sum \limits_j \bm d_{ \Omega_j}^{R,A}(\varepsilon)e^{i \Omega_j t},~~~~~~~~~~~~~
  \label{solution_first}
  \end{eqnarray}
    with
  \begin{eqnarray}
  \bm d_{\Omega}^{R,A}(\varepsilon) = K_d^{R,A}(\varepsilon, \Omega) [(\bm E_\Omega \times \bm n) \times \bm h], 
  \label{d_omega}
  \end{eqnarray}
  which is our second main result. The kernel in Eq.~(\ref{d_omega}) in the limit  $\xi_F \ll d_F \ll \xi_{\Omega_i}$ takes the form:
     \begin{align}  \label{Eq:Kernel}
     K_d^{R,A} (\varepsilon, \Omega ) =  \pm \frac{4e \tilde \alpha \xi_F^2
     (\gamma \xi_F)|\Delta|\sin (\chi/2) 
     \hat \tau_2}{d_F \bigl[ (\varepsilon \pm i \delta)^2 - (\Omega/2)^2 \bigr] \Omega}.
 \end{align}
  
 \begin{figure}
 \centerline{$
 \begin{array}{c}
 \includegraphics[width=3.5in]
 {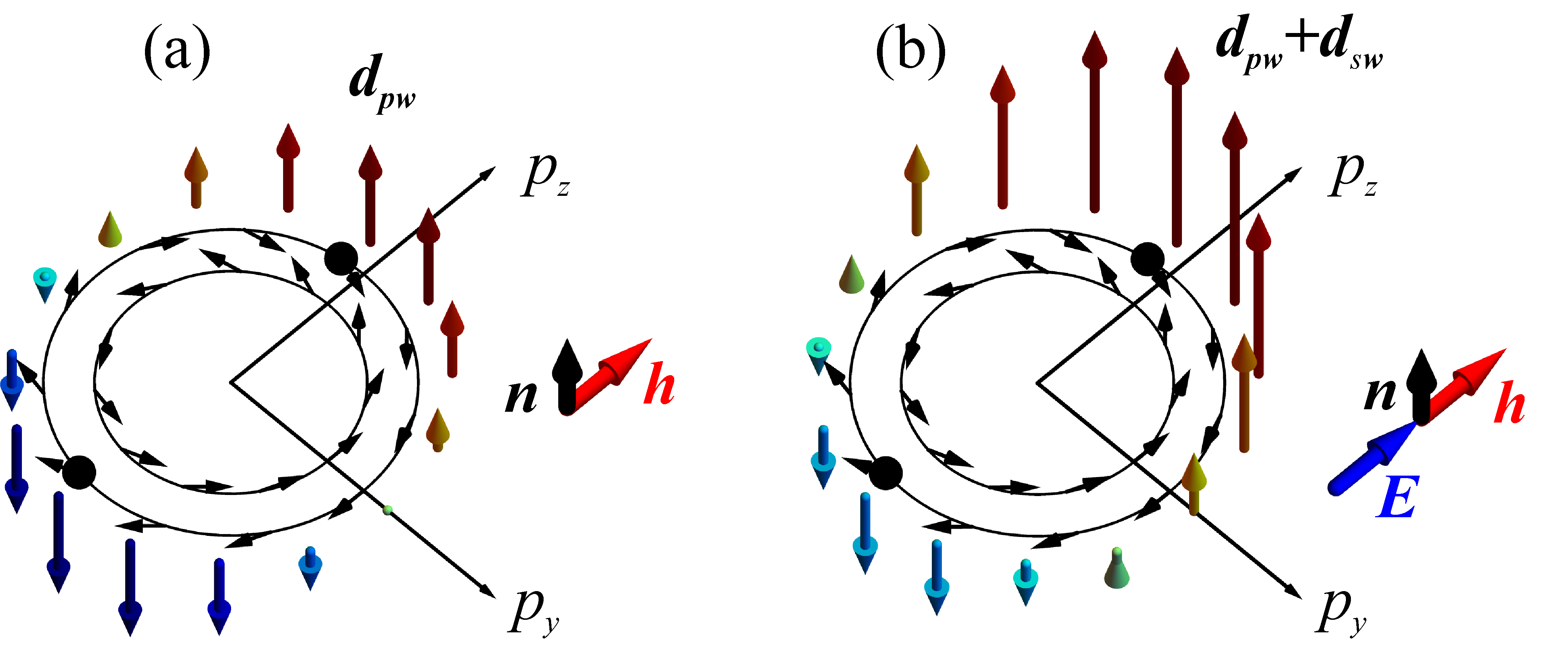} 
 \end{array}$}
 \caption{\label{Fig:2} 
  {\bf Mechanism behind the formation of spin-triplet superconducting correlations.}
  Helical Fermi surface cross-sections $p_x=0$
    in the region close to S/F interface with Rashba SOC and exchange field.  
    Local directions of spin quantization axes are marked by black arrows. Small black spheres show
  states with opposite momenta $\bm p$ and   $-\bm p$.
  Rashba SOC vector is $\bm n = \bm x$ and exchange field $\bm h = h \bm z$. 
  (a) Non-collinearity of spins at $\bm p$ and $-\bm p$ results in the spin-triplet pairing with p-wave spin vector 
  $\bm d_{pw} \propto ( \bm n \times \bm p)\times \bm h  \parallel \bm x$.
   (b) Alternating electric field $\bm E$ shown by the blue arrow results in the mixing
   between p and s-wave pairing amplitudes.   }
 \end{figure}

The qualitative physical picture of the LRTs generation in Eq.~(\ref{bc_d}) is determined by the "hidden" p-wave correlations which are induced by the interplay of SOC and exchange field $\bm h$ at the S/F interface and the p-wave to s-wave conversion induced by the electric field drive.  
First, we
recall that the spin-splitting by the exchange field near the S/F interface\cite{bergeret2005odd, Buzdin2005,Linder2015,Eschrig2015}
provides the spin-mixing and thus induces the s-wave spin-triplet superconducting correlations   $\bm d_{sw}^{short} = (i \gamma \xi_F F_{bcs}/2h)\bm h$ with zero spin projections $S_z =0$ on the quantization axis $\parallel \bm h$,  where $F_{bcs}^{R,A} = \mp \Delta/(\varepsilon \pm i \delta)$.  These correlations are short-ranged in the ferromagnet.
The SOC does not provide spin splitting  (neglecting the small terms of the order of $\alpha/v_F$, where $v_F$ is the Fermi velocity), but it induces the 
momentum-dependent  rotation of the spin quantization axis $\bm h \to \bm h + \alpha \bm n\times \bm p$  where $\bm p$ is the electron momentum. This provides a
 conversion of the spin-triplet s-wave  $\bm d_{sw}^{short}$ to the  spin-triplet p-wave $\bm d_{pw}$ correlations. 
 This conversion follows from the standard quasiclassical Eilenberger equation in the presence of SOC. In the diffusive limit it yields a general local relation 
$ \bm d_{pw}  = 
  i (D \alpha/p)   
 ( \bm p\times \bm n) \times \bm d_{sw}^{short}$, see Fig.\ref{Fig:2}a.
Similar mechanisms of the triplet p-wave component generation take plays in various topological superconductivity platforms.
 
   Thus spin-triplet correlations are characterized by the spin vector
  $\bm d_{pw} (\bm p, \varepsilon) = F_{pw}(\varepsilon) \bm h\times (\bm n\times {\bm p}) $ 
with the amplitude $F_{pw}(\varepsilon) = i\alpha  \gamma D \xi_F F_{bcs}/2h p$.
   The p-wave pairing exists only in the surface layer with non-zero SOC $\alpha(x)\neq 0 $. Outside this layer it vanishes at the mean free path length and therefore do not penetrate into the ferromagnet. However,  the electric field induces the condensate momentum $\bm p_s = -2 i e \bm E_\Omega /\Omega$ providing coupling between orbital p-wave and s-wave components via the added energy Doppler shift  
 \cite{PhysRevB.7.1001,kohen2006probing} $\bm p\cdot \bm p_s /m$. 
 As a result the  amplitude of triplet correlations is given by  $F_{pw}(\varepsilon + \bm p\cdot\bm p_s/m) \approx F_{pw}(\varepsilon) + (\bm p\cdot\bm p_s/m) \partial_\varepsilon F_{pw}$.
  This modification of the pairing amplitude produces the additional s-wave component of the spin vector  
$\bm d_{sw}^{long} \propto \bm h\times (\bm n\times \bm E)$ with spin projections $S_z = \pm 1$ on the quantization axis, which is  suppressed neither by the impurity scattering nor by the exchange field.
The vector field $\bm d= \bm d_{pw} + \bm d_{sw}^{long}$ is shown schematically in Fig.~\ref{Fig:2}b. 
As a result we obtain the conversion of  $s$-wave $S_z=0$ to the $s$-wave $S_z=\pm 1$ pairs through the local p-wave correlations and the Doppler shift. On the level of Usadel equations, which only operate with $s$-wave Green's functions, this three-stage process results in the non-zero r.h.s. of the boundary conditions (\ref{bc_d}).   


{\it Photo-induced Josephson current.}
  The overlapping between two LRT amplitudes penetrating from the both S/F interfaces gives rise to the nonzero Josephson effect. For the case of a harmonic electromagnetic wave we get the current-phase relation:
     \begin{align} \label{Eq:CurrentJJ}
     I (\chi,t)= [I_{dc}^c + I_{2\Omega}^c \cos(2\Omega t) ]\sin\chi .  
   \end{align}
      Note that here both the dc and double-frequency critical current amplitudes are determined by the alternating  electric field $I_{dc}^c \propto E_\Omega E_{-\Omega}$ and $I_{2\Omega}^c \propto  E_{\Omega}^2$.  The particular values of the critical currents $I_{dc}^c,\; I_{2\Omega}^c $ can be found in the supplementary material \cite{supplementary}. 
  By the order of magnitude $I_{dc}^c,\; I_{2\Omega}^c \sim I_0 $ , where 
  %
    \begin{align}\label{Eq:Ic}
    I_0 = - \sigma_F S (\Delta/ed_F) (2\tilde\alpha\gamma\xi_F/\pi)^2   (\Delta /T)^2  P/P_c
  \end{align}
  where $S$ is the junction area, 
 $P = c |E_\Omega|^2$ 
is the radiation power,  
 $P_c = (c\hbar /e^2) \hbar \Omega^2/\xi_S^2 $
is the radiation power needed to speed up the Cooper pairs to the depairing velocity.
  The scale 
 $I_0$
 can be estimated using the typical   parameters of JJ with ferromagnetic interlayers\cite{fominov2007josephson}: the junction area is $50 \times 50 \;\mu{\rm m}^2$,
 $\sigma_F\sim (50\;  \mu\Omega\; {\rm cm} )^{-1}$, 
 $d_F\sim \gamma^{-1} \sim 5 \xi_F$ and 
 $D\sim 10\; {\rm cm^2/s}$, 
 $h\sim 500\; {\rm K}$ so that
 $\xi_F \sim 3 {\rm nm}$. 
 For the superconducting gap in Nb $\Delta \sim 10 {\rm K}$ so that the critical current is 
 $I_0 \sim 10^{-1}(p_s\xi_S)^2 \tilde \alpha^2$ A. Taking $\tilde \alpha \sim 0.1-1$ 
 \cite{lo2014spin,banerjee2018controlling, ast2007giant,Triola2016,cayao2018odd} 
    we get the current $I_0/(p_s\xi_S)^2 \sim 10^{-1} - 10^{-3}$ A. 
  Given that $\xi_S \approx 30$ nm we get 
$P_c \approx 10 (\Omega/GHz)^2$  W /m$^2$. 
Therefore such a JJ is quite sensitive to the radio-frequency and microwave irradiation. A typical cell phone at one 
meter distance generates microwave
radiation with $\Omega \approx 3-4$ GHz and $P\sim P_c$ which 
induces rather large  currents  $I_0 \sim 10^{-1} - 10^{-3}$ A. At the same time the power sensitivity strongly decreases with the frequency rise.  
For the frequency corresponding to the cosmic background radiation 
$P_c \approx 10^6 $ W /m$^2$  so that the power density $P= 10^{-5}$
W /m$^2$ induces rather small critical current $I_0 \sim 10^{-12} - 10^{-15}$ A.  
Still, it is possible to induce large critical current using
THz and visible light radiation sources. 
The $1$ THz radiation with power $1$ mW /mm$^2$ yields $I_0 \sim 10^{-5} -10^{-7}$ A. 
Laser beam of the frequency about $\Omega \sim 10^{6}$ GHz 
carrying the power 
$1$ mW focused into the spot of $1$ $\mu$m$^2$ size induces
the critical current $I_0 \sim 10^{-6} - 10^{-8}$ A which is well within the measurable limits. 

\begin{figure}[h!]
 \centerline{$
 \begin{array}{c}
 \includegraphics[width=3.5in]{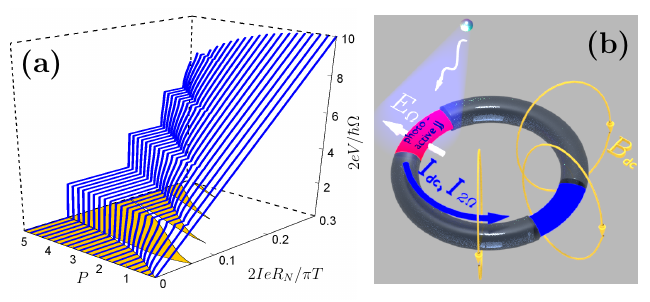} 
 \end{array}$}
 \caption{\label{Fig:shapiro} 
 (a)
  {  IV-characteristics of the irradiated photo-active JJ at a constant applied current $I$.} Different blue curves correspond to different values of the applied radiation power $P=2eR_N I_{dc}^c/(\hbar \Omega)$ at a given frequency, $R_N$ is the JJ resistance in the normal state. $\hbar \Omega/\pi T = 0.03$.  Shown with yellow shadings are the domains in $(I,P)$ plane with constant voltage generated across the JJ. 
  (b)
   { Schematic picture of the photo-magnetic SQUID}. The device consists of the photo-active Josephson junction  (red weak link) and the usual JJ (blue weak link). 
   Electric field  $\bm E_\Omega e^{i\Omega t}$ from the incoming radiation switches on both dc $I_{dc}$ and $I_{2\Omega}e^{2i\Omega t}$ components of the circulating current. The dc component produces spontaneous magnetic field $\bm B_{dc}$. 
}
 \end{figure}

In IV-characteristics of conventional Josephson junctions Shapiro steps can be observed at $2eV = \hbar n \Omega$ under periodic external perturbations: a periodic applied current or under irradiation. For the system under consideration the driving electric field is parallel to the interfaces and thus  does not induce the voltage across the junction. This geometry is qualitatively different from the usual experiments on microwave-induced Shapiro steps. Nevetheless, the presence of the second harmonic contribution to the critical Josephson current (\ref{Eq:CurrentJJ}) leads to the Shapiro steps with unusual properties even under a constant applied current. For the system under consideration the Shapiro steps take place at voltages $2eV = 2 \hbar n \Omega$. That is only the even-numbered Shapiro steps show up in the photo-active JJ. Obviously, it is a consequence of the fact that the ac component of the current oscillates with a frequency twice larger than the externally applied source. The second essential difference with the conventional case is that the value of the critical current grows with the radiation power, as it is demonstrated by different curves in Fig.~\ref{Fig:shapiro}(a). This is a signature of the irradiation-induced LRT correlations.  
  

 {\it  Josephson photo-magnetic devices }
  Electric-field induced current across the JJ, described by Eqs.(\ref{Eq:CurrentJJ}), (\ref{Eq:Ic}), provides an interesting possibility to create photo-magnetic devices based on the superconducting loops with the weak links formed by the radiation-controlled JJ.  
  We show that applying the radiation as it is shown in the schematic Fig.\ref{Fig:shapiro}(b),   it is possible to generate spontaneous currents circulating in the loop, which in turn produce a dc component of the magnetic field $\bm B_{dc}$.  
 
  We consider a dc SQUID with  one of the branches connected by photo-active JJ and the other by a $\pi$-JJ which is used as a passive phase shifter element as in the "quiet" superconducting qubits proposals \cite{ioffe1999environmentally,PhysRevB.63.174511,feofanov2010implementation} and rapid single flux quantum logic devices \cite{ustinov2003rapid}.  Dynamics of Josephson phases $\chi_1$ across the photo-active JJ  and $\chi_2$ across the $\pi$-JJ is determined by the system of coupled sine-Gordon equations \cite{barone1982physics}, which is similar to the one used for the standard dc SQUID. The essentially different effects are determined by the two factors. The first one is the possibility of various parametric effects
due to the time-dependent current amplitude in the photo-active JJ. These effects can be expected for the frequencies comparable with the eigen frequency of the superconducting loop $\omega_0 = 1/\sqrt{LC}$. 
Here we consider the opposite case when $\Omega\ll \omega_0$ and use the second non-trivial property of the system, which is the critical current of the photo-active JJ controllable by the radiation power.  In this case we can separate the time scales corresponding to rapid oscillations and slow period-averages drift described by the coordinate 
  ${\bm\chi}=  (\bar \chi_1, \bar \chi_2)$  where $ \bar \chi_k =  \Omega \int_0^{\Omega^{-1}} \chi_k d t$. 
  In the absence of external irradiation 
 there are no currents and phase differences are 
 $\bar\chi_{1,2} =0$. 
  Radiation switches on the photo-active JJ.  
  Then gradually increasing  the radiation power we get that the zero-current state becomes unstable  under the following condition \cite{supplementary} \begin{align} \label{Eq:Instability}
 I_{dc}^c > \frac{\Phi_0}{2\pi} \frac{\omega_0\omega_p}{\sqrt{\omega_0^2 + \omega_p^2}}
 \end{align}
 where $\omega_p = \sqrt { 2\pi I_\pi^c /C\Phi_0 }$ is the plasma frequency corresponding to the  $\pi$-JJ.
  In case of the typical values $\omega_p=\omega_0 \sim 10$ GHz
 we get the threshold value in the r.h.s. of Eq.~(\ref{Eq:Instability}) about   
 $10^{-6}$ A.
 
 Once the condition (\ref{Eq:Instability}) is satisfied the SQUID switches to the state with spontaneous dc current $I_{dc}$ and constant magnetic field $\bm B_{dc}$ shown schematically in Fig.\ref{Fig:shapiro}(b). 
  The photo-induced magnetic flux
 magnitude can be estimated as $L I_{dc}^c$. For the typical values of the SQUID loop inductance $L \approx 10^{-11}$ H and $I_{dc}^c \approx 10^{-6}$ A we get the flux of $10^{-2} \Phi_0 $. 
 
One can obtain the photo-magnetic response  without any threshold for the incoming power provided the second branch of the SQUID contains the  Josephson phase battery \cite{Buzdin2008,szombati2016josephson,assouline2019spin,strambini2020josephson} 
based on the JJ with shifted current-phase relation 
 $I = I_{\varphi}^c \sin (\chi - \varphi_0)$ with $\varphi_0 \neq \pi n$.
 Such photo-magnetic element generates dc current 
 $I_{dc} \approx I_{dc}^c \cos\varphi_0$ and the corresponding magnetic field 
 $\bm B_{dc}$ being exposed to any arbitrary small radiation power.  

 In conclusion, we have demonstrated the
  possibility of generating
    dynamic spin-triplet superconducting order which emerges under non-equilibrium conditions induced by the   
   alternating electric field.  
 Qualitatively the obtained effect arises due to the partial conversion of the $p$-wave triplet superconductivity, taking place in the presence of the Rashba SOC and ferromagnetism, to the $s$-wave odd-frequency triplet correlations. 
The conversion is caused by the Doppler shift of the quasiparticle spectrum induced by the non-stationary condensate motion under the action of the electric field. 
The detailed qualitative discussion  and
development of the microscopic model of this mechanism are provided. 
We  
propose a scheme of a Josephson transistor which can be switched by the ac current and a photo-magnetic SQUID, which generates magnetic fields under irradiation.

\begin{acknowledgments}
The work of M.A.S was supported by the Academy of Finland (Project No. 297439) and  Russian Science Foundation, Grant No. 20-12-00053.
The work of I.V.B and A.M.B has been carried out within the state task of ISSP RAS with the support by RFBR grants 19-02-00466, 18-52-45011 and 18-02-00318. I.V.B. also acknowledges the financial support by Foundation for the Advancement of Theoretical Physics and Mathematics “BASIS”.
\end{acknowledgments}

\section{Supplemental Material}

\subsection{Formalism: nonstationary Usadel equations in the mixed representation and boundary conditions}
 Quantitatively the action of the external electric  field
on the SFS junction  
can be described by the Usadel equation for the quasiclassical Green's function $\check g(\varepsilon, t, \bm r)$.
In the mixed $(\varepsilon, t)$ representation it  takes the form:
  \begin{eqnarray}
  iD \hat\partial_{\bf r}
  \bigl( \check g \otimes 
  \hat\partial_{\bf r} \check g \bigr) 
  = 
  \bigl[\hat \tau_3 (\varepsilon - \bm h \hat {\bm \sigma} + \hat \Delta ), 
  \check g  \bigr]_\otimes,
  \label{usadel_mixed}
  \end{eqnarray}
  where $\hat \tau_i$ and $\hat \sigma_i$ are Pauli matrices in particle-hole and spin spaces, respectively, and $\hat {\bm \sigma} = (\hat \sigma_1, \hat \sigma_2, \hat \sigma_3)^T$. The Green's function $\check g(\varepsilon, t, \bm r)$ is a $8 \times 8$ matrix in the direct product of particle-hole, spin and Keldysh spaces. It describes two-particle correlations and depends on "center of mass"  coordinate $\bm r$. In addition the mixed representation means the Fourier transform $t_1-t_2 \to \varepsilon$, but the resulting Green's function still depends on time via $t=(t_1+t_2)/2$ because of the non-stationary character of the problem under consideration. In the mixed representation usual multiplication is replaced by the $\otimes$-product, which is defined as $\check A \otimes \check B = \exp \bigl( (i/2)(\partial_{\varepsilon_A} \partial_{t_B} - \partial_{\varepsilon_B} \partial_{t_A}) \bigr]\check A(\varepsilon, t) \check B(\varepsilon, t)$.  In case $\bm A(t) = \bm A_\Omega \exp[i \Omega t]$ the $\otimes$-products are reduced to $\bm A(t) \otimes \check g(\varepsilon, t) = \bm A_{\Omega} \check g(\varepsilon+\Omega/2,t)\exp[i \Omega t]$ and $\check g(\varepsilon, t) \otimes \bm A(t)= \bm A_{\Omega} \check g(\varepsilon-\Omega/2,t)\exp[i \Omega t]$.
  
  In Eq.~(\ref{usadel_mixed}) $D$ is the diffusion constant and $\hat \Delta = |\Delta(x)|exp[i \chi (x) \hat \tau_3]\hat \tau_1$. We assume $|\Delta(x)| = 0$ in the interlayer of the Josephson junction $-d_F/2 \leq x \leq d_F/2$, while $|\Delta(x)| = \Delta $ and $\chi(x) = \mp \chi/2$ in the left (right) superconducting leads. 
 The differential superoperator in Eq.~(\ref{usadel_mixed}) is
 
  \begin{align}\label{Eq:GradGen}
 \hat\partial_k \check g= 
 \nabla_k \check g  - 
 i\bigl[\alpha \hat {\cal A}_k + \frac{e}{c}  A_k \hat\tau_3, \check g \bigr]_\otimes
 \end{align}
  The presence of an alternating electric field $\bm E (t)$ is described by the 
 time-dependent vector potential 
 $ \bm E = -  \partial_t \bm A /c $. 
  Rashba-type SOC in (\ref{Eq:GradGen}) is described by the SU(2) gauge field 
 $\hat {\cal A}  = \bm n \times 
 \hat{\bm \sigma} $. 
  
  In the F region the superconducting correlations are small so that we linearize the Usadel equation by assuming $\hat g^{R,A} = \pm \hat\tau_3 + \check f^{R,A}$ where the anomalous part $\check f^{R,A}= \hat f_s^{R,A} + \hat { \bm f}_t^{R,A}\bm\sigma$ can be written as the sum of spin-singlet $f_s^{R,A}$ and spin-triplet $\hat {\bm f}_t^{R,A}$ components. Then we obtain the linearized equation 
  \begin{eqnarray}
  2 \varepsilon \check f^{R,A} - \{ \bm h \hat {\bm \sigma}, \check f^{R,A} \} \mp 2i \hat \tau_3 \hat \Delta = \nonumber \\
  \pm i D \Bigl( \partial_x^2 \check f^{R,A} - \frac{e^2}{c^2}\sum \limits_{ij}\bm A_{\Omega_i} \bm A_{\Omega_j} e^{i(\Omega_i+\Omega_j)t}\times  \nonumber \\
  \sum \limits_{a,b = \pm 1} \check f^{R,A}(\varepsilon + \frac{a \Omega_i + b \Omega_j}{2}) -\frac{2 \alpha e \hat \tau_3}{c}\sum \limits_i \bm A_{\Omega_i} e^{i \Omega_i t} \times \nonumber \\
  \bigl[ \bm n \times \hat {\bm \sigma}, \bigl( \check f^{R,A}(\varepsilon + \frac{\Omega_i}{2})+\check f^{R,A}(\varepsilon - \frac{\Omega_i}{2}) \bigr) \bigr] - \nonumber \\
  \alpha^2  \bigl[ \bm n \times \hat {\bm \sigma}, [\bm n \times \hat {\bm \sigma},\check f^{R,A}] \bigr]\Bigr)~~~~~~~~~~~~~ 
  \label{Usadel_linearized}
  \end{eqnarray}
   We assume that the SOC is localized near FS interfaces at the lengthscale $d_{so}$ much smaller that $\xi_F$, and also we assume the realistic limit $\xi_F \ll d_F \ll \xi_{\Omega_i}=\sqrt{D/\Omega_i}$. In this case the SO coupling is absent in the most part of the interlayer and can be taken into account as a boundary condition condition at $x = \mp d_F/2$ and the Usadel equation Eq.~(\ref{Usadel_linearized}) in the interlayer region takes the usual form:
   \begin{eqnarray}
   \pm i D \partial_x^2 \check f^{R,A} = 2 \varepsilon \check f^{R,A}- \{ \bm h \hat {\bm \sigma}, \check f^{R,A} \}. 
   \label{Usadel_linearized_1}
   \end{eqnarray}
   Further we assume that the surface SOC strength $\tilde{\alpha} = \int dx \alpha (x) \ll 1$ and also $p_s \xi_S \ll 1$, $p_s \xi_{\Omega_i} \ll 1$ and perform calculations in the framework of the perturbation theory with respect to these parameters. Therefore, the small terms $\propto \bm A_{\Omega_i} \bm A_{\Omega_j}$, which are not spin-active and do not lead to any singlet-triplet or SRT-LRT conversion, are neglected.
   
   In order to obtain boundary condition we should integrate Eq.~(\ref{Usadel_linearized}) over the regions $-d_F/2<x<-d_F/2+d_{so}$  and $d_F/2-d_{so}<x<d_F/2$ with nonzero SO coupling near the S/F interfaces. The result takes the form:
   \begin{eqnarray}
   n_x \partial_x \check f_t^{R,A} = \frac{2e}{c} \tilde \alpha \hat \tau_3 \sum \limits_i \bm A_{\Omega_i} e^{i \Omega_i t} \bigl[ \bm n \times \hat{\bm \sigma}, \nonumber \\
   \bigl( \check f_t^{R,A}(\varepsilon + \frac{\Omega_i}{2})+\check f_t^{R,A}(\varepsilon - \frac{\Omega_i}{2}) \bigr) \bigr],
   \label{bc}
   \end{eqnarray}
  where $n_x = \pm 1$ for the left(right) S/F interface. There are some terms, which are present in Eq.~(\ref{Usadel_linearized}), but are omitted  in Eq.~(\ref{bc}). The reason is that they do not lead to the SRT $\to$ LRT conversion and, therefore, are small in the framework of the perturbation theory. 
  
  The explicit spin structure of the triplet correlations is revealed via the $\bm d$-vector as $\check f_t^{R,A} = \bm d^{R,A} \hat {\bm \sigma}$. In terms of the $\bm d$-vector the boundary condition Eq.~(\ref{bc}) can be rewritten as
  \begin{eqnarray}
  n_x \partial_x \bm d^{R,A} = \frac{4ie}{c} \tilde \alpha \hat \tau_3 \sum \limits_i e^{i \Omega_i t} \bigl( \bm A_{\Omega_i} \times \bm n \bigr) \times \nonumber \\
  \bigl[ \bm d^{R,A}(\varepsilon + \frac{\Omega_i}{2})+\bm d^{R,A}(\varepsilon - \frac{\Omega_i}{2}) \bigr].
   \label{bc_d}
  \end{eqnarray}
  
  The boundary condition for the singlet component is the usual Kupriyanov-Lukichev boundary condition
  \begin{eqnarray}
   n_x \nabla_x \hat f_s^{R,A} =  \gamma \hat F_{bcs}^{R,A},
      \label{bc_singlet}
  \end{eqnarray}
  where $\hat F_{bcs}^{R,A} = \mp \tau_3 \hat \Delta /(\varepsilon \pm i \delta)$.
  
  \subsection{Details of the solution for the LRTC}
  
  In the considered limit  $\xi_F\ll d_F \ll \xi_{\Omega_i}$ solution of system Eqs.~(\ref{Usadel_linearized_1}),(\ref{bc}) and (\ref{bc_singlet}) up to the first order with respect to the parameter $\tilde \alpha p_s \xi_S$ reads as follows:
  \begin{eqnarray}
  \check f^{R,A} = \check f_s^{R,A} + \bm d_h^{R,A}\hat {\bm \sigma} + \bm d^{R,A}\hat {\bm \sigma}. \label{solution_sum}
  \end{eqnarray}
  The short-range component of the anomalous Green's function $f_s^{R,A}$ and $\bm d_h^{R,A}$ can be found at each of the interfaces separately because they decay at the distance $\sim \xi_F$ from the interface. Up to the zero order with respect to $\tilde \alpha p_s \xi_S$ directly at the left(right) interface they take the form:
  \begin{eqnarray}
  \left(
  \begin{array}{c}
  \check f_s^{R,A} \\
  \bm d_{h,l(r)}^{R,A}
  \end{array}
  \right)= -\frac{\gamma \xi_F}{2}\hat F_{bcs,l(r)}^{R,A} 
  \left(
  \begin{array}{c}
  1 \\ 
  \mp i
  \end{array}
  \right), 
  \label{solution_short}
  \end{eqnarray}
  where $\hat F_{bcs,l(r)}^{R,A} = \mp \tau_3 \hat \Delta_{l,r} /(\varepsilon \pm i \delta)$ are BCS anomalous Green's functions in the left (right) superconductor.
  \begin{eqnarray}
  \bm d^{R,A} (\varepsilon,t) = \sum \limits_j \bm d_{ \Omega_j}^{R,A}(\varepsilon)e^{i \Omega_j t},~~~~~~~~~~~~~
  \label{solution_first}
  \end{eqnarray}
    with
  \begin{eqnarray}
  \bm d_{\Omega}^{R,A}(\varepsilon) = \pm \frac{2e}{c}\frac{\tilde \alpha D}{d(\varepsilon\pm i \delta)}\hat \tau_3 \bigl( \bm A_{\Omega} \times \bm n_r \bigr)\times \nonumber \\
  \sum \limits_{a = \pm 1}\Bigl[ \bm d_{h,r}^{R,A}(\varepsilon+a\Omega/2) - \bm d_{h,l}^{R,A}(\varepsilon+a\Omega/2) \Bigr]
  \label{dperp_first}
  \end{eqnarray}
  Substituting $\bm d_{h,l(r)}^{R,A}$ from Eq.~(\ref{solution_short}) into Eq.~(\ref{dperp_first}) we obtain $\bm d_{\Omega}^{R,A}(\varepsilon) = K_d^{R,A}(\varepsilon, \Omega) [(\bm E_\Omega \times \bm n) \times \bm h]$ with the kernel expressed by Eq.~(8) of the main text.
  
  \subsection*{Details of calculation of the Josephson current}
  
  The expression for the Josephson current via the quasiclassical Green's function reads as follows:
  \begin{equation}
  I = -\frac{\sigma_F S}{16e} \int d \varepsilon {\rm Tr}_4 \Bigl[ \hat \tau_3 \check g \otimes \hat {\bm \nabla}\check g \Bigr]^K,   
  \end{equation}
  where $\sigma_F$ is the conductivity of the ferromagnet and $S$ is the junction area. It is convenient to calculate the Josephson current at one of the interfaces. Then making use of the Kuprianov-Lukichev boundary conditions it can be expressed via the  Green's function in the ferromagnet near the interface $\check g_{l(r)}$ and the Green's function in the leads $\check g_{s,l(r)}$:
  \begin{equation}
  I = \mp \frac{\sigma_F \gamma S}{32e} \int d \varepsilon {\rm Tr}_4 \Bigl\{ \hat \tau_3 \bigl[\check g_{BCS,l(r)}, \check g_{l(r)} \bigr]_\otimes^K\Bigr\}.
  \label{current_interface}
  \end{equation}
    By linearizing the expression with respect to the anomalous Green's functions and assuming the equilibrium quasiparticle distribution Eq.~(\ref{current_interface}) can be further rewritten as follows
  \begin{eqnarray}
  I = \mp \frac{\sigma_F \gamma S}{16e} \int d \varepsilon {\rm Tr}_2 \Bigl[ \left\{ \frac{\hat \Delta_{l(r)}}{\varepsilon+i\delta}, \hat f_{s,l(r)}^{R}\otimes \tanh \frac{\varepsilon}{2T} \right\}_\otimes + \nonumber \\
  \left\{ \frac{\hat \Delta_{l(r)}}{\varepsilon-i\delta}, \tanh \frac{\varepsilon}{2T} \otimes \hat f_{s,l(r)}^{A}  \right\}_\otimes\Bigr].~~~~~~~~~~~~~~
  \label{current_interface_lin}
  \end{eqnarray}
  
  The singlet part of the anomalous Green's function entering this equation should be taken up to the second order with respect to $\tilde \alpha p_s \xi_S$ because in the zero order  the overlapping of $f_s$ coming from the both interfaces is negligible in the interlayer and, therefore, it does not contribute to the Josephson current. The second order correction to $f_s$ can be found from Eq.~(\ref{Usadel_linearized_1}) and boundary conditions Eqs.~(\ref{bc_d})-(\ref{bc_singlet}) and takes the form:
  \begin{eqnarray}
  \hat f_{s,l(r)}^{(2)R,A} = \mp \frac{2e}{c}\tilde \alpha \xi_F \hat \tau_3 \sum \limits_i e^{i\Omega_i t } \times \nonumber \\
  \Bigl( \bigl[ (\bm A_{\Omega_i}\times \bm n_{l(r)})\times \sum \limits_{a=\pm 1}\bm d^{R,A} (\varepsilon+\frac{a\Omega_i}{2}) \bigr] \bm n_h \Bigr),
  \label{singlet_2}
  \end{eqnarray}
  where $\bm n_h = \bm h/h$.
  
  In case if the long-range effect is caused by a monochromatic electromagnetic wave the amplitude of the Josephson current has the dc component $\propto E_\Omega E_{-\Omega} $ and
 the second harmonic one
 $\propto E_\Omega^2$. At first let us consider the dc component of the Josephson current:
 \begin{eqnarray}
 \hat f_{s,dc}^{(2)R,A} = \mp \frac{2e}{c}\tilde \alpha \xi_F \hat \tau_3 \Bigl\{\bigl[ (\bm A_{\Omega}\times \bm n)\times \nonumber \\
 \sum \limits_{a=\pm 1}\bm d_{-\Omega}^{R,A} (\varepsilon+\frac{a\Omega}{2}) \bigr] \bm n_h 
 + \Omega \to -\Omega \Bigr\}
 \label{fs_dc}
 \end{eqnarray}
  Although the second order contribution to $f_s$ decays quite rapidly from the interface (at the distance of the order of $\xi_F$), $f_s^{(2)}$ at the left (right) interface contains information about the superconducting correlations at the opposite interface and, therefore, contribute the the Josephson current. This information is transferred via the LRT component $\bm d_{\Omega}$, which decays much slower and is approximately constant inside the interlayer in the considered regime $d_F \ll \xi_\Omega$. Substituting the LRT component $\bm d_{\Omega}$ into Eq.~(\ref{fs_dc}) from Eq.~(\ref{dperp_first}), we finally obtain the dc Josephson current from Eq.~(\ref{current_interface_lin}):
  \begin{eqnarray}
  I_{dc} = I_{dc}^c \sin \chi,~~~~~~~~~~~~~~~~~ \nonumber \\ 
  I_{dc}^c =\frac{4 I_0 |\cos \Phi|^2}{a^2} \Bigl[{\rm Re} [\psi(\frac{1+ia}{2})]-\psi(\frac{1}{2}) \Bigr],
    \label{current_dc}
  \end{eqnarray}
  where $a = \Omega/\pi T$, $\psi(x)$ is the digamma function and we have introduced $\cos \Phi = (\bm E_\Omega \bm n_h)/|\bm E_\Omega|$. $I_0$ is defined in Eq.~(10) of the main text and is estimated there by the order of magnitude for different types of radiation. 
  
  Now let us consider the second harmonic contribution to the Josephson current. For the second harmonic contribution to the anomalous Green's function we obtain:
  \begin{eqnarray}
 \hat f_{s,2\Omega}^{(2)R,A} = \mp\frac{2e}{c}\tilde \alpha \xi_F \hat \tau_3 \bigl[ (\bm A_{\Omega}\times \bm n)\times \nonumber \\
 \sum \limits_{a=\pm 1}\bm d_{\Omega}^{R,A} (\varepsilon+\frac{a\Omega}{2}) \bigr] \bm n_h 
 \label{fs_2omega}
  \end{eqnarray}
 Substitution of the LRT component $\bm d_\Omega$ from Eq.~(\ref{dperp_first}) leads us to the following answer for the second harmonic current contribution:
  \begin{eqnarray}
  I_{2\Omega} = I_{2\Omega}^c \cos (2\Omega t ) \sin \chi, \nonumber \\
  I_{2\Omega}^c = \frac{2 I_0 |\cos \Phi|^2}{a} \times \nonumber \\
  \Bigl| \sum \limits_{n=0}^\infty  \Bigl( \frac{1}{(2n+1 + 2 i a)^2}-\frac{1}{(2n+1)^2} \Bigr)\Bigr|.
  \label{current_2omega}
  \end{eqnarray}
  
    \begin{figure}[h!]
 \centerline{$
 \begin{array}{c}
 \includegraphics[width=2.6in]{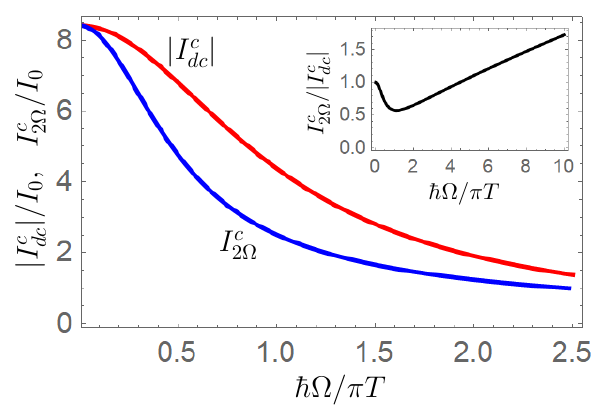} 
 \end{array}$}
 \caption{\label{Fig:harmonics} 
  {\bf DC and double-frequency components of the critical current.}
   $I_{dc}^c$(red) and $I_{2\Omega}^c$(blue) as functions of $\Omega/\pi T$. The currents are normalized to $I_0$. Insert demonstrates the ratio $I_{2\Omega}^c/I_{dc}^c$ as a function of frequency.
     }
 \end{figure}
 
 The behavior of  dc $I_{dc}^c$ and second harmonic $I_{2\Omega}^{c}$ components of the current as functions of the frequency  $\Omega/\pi T$ is plotted in Fig.~\ref{Fig:harmonics}.
 It is seen that the dc component dominates at $\hbar \Omega/\pi T \lesssim 1$, what corresponds to the frequencies range up to THz range at not very low cryogenic temperatures.
 
\subsection{Josephson photo-magnetic device}

We assume that the photo-induced  currents are significantly small so that   
  $ \omega_{ph} \ll \Omega$,  where $\omega_{ph} = \sqrt{ 2\pi I_{dc}^c / C\Phi_0}$
  is the corresponding Josephson plasma frequency. Under this condition we can neglect the contribution to the force of the order $(\omega_{ph}/\Omega)^2$ coming from averaging the second-harmonic contribution to the Josephson current.
    Then the period-averaged phases  satisfy equation 
 \begin{align} \label{Eq:KapitzaAverage}
 & \partial_t^2 {\bm\chi} + \gamma_{RC} \partial_t {\bm\chi} 
 = - \partial_{\bm\chi} 
 U ({\bm\chi})   
 \\
 & U ({\bm\chi}) = \frac{\omega_0^2}{2} (\chi_1 + \chi_2)^2 +   
 \omega_p^2 \cos \bar\chi_2 -
 \omega_{ph}^2 \cos \bar\chi_1
 \end{align}
 where $\omega_0 = 1/\sqrt{LC}$ and 
 $\gamma_{RC} = 1/RC$, which are the resonant frequency and decay rate, respectively. $\omega_p = \sqrt { 2\pi I_\pi^c /C\Phi_0 }$ is the plasma frequency of the $\pi$-JJ.
 
 In the absence of external irradiation $\omega_{ph}=0$, so 
 there are no currents and phase differences are 
 $\bar\chi_{1,2} =0$. 
  Radiation switches on the photo-active JJ $\omega_{ph} \neq 0$ 
  and increasing gradually the radiation power we get that the zero-current state becomes unstable  under the following condition 

 \begin{align} \label{Eq:Instability}
 I_{dc}^c > \frac{\Phi_0}{2\pi} \frac{\omega_0\omega_p}{\sqrt{\omega_0^2 + \omega_p^2}}.
 \end{align}

\bibliography{refs2}

\end{document}